\documentclass[aip, apl, reprint, floatfix]{revtex4-1}
\usepackage{amsfonts, amssymb, amsbsy, amsmath}
\usepackage{graphicx}
\usepackage{hyperref}
\usepackage{bm}
\usepackage{dcolumn}

\newcommand{\equref}[1]{(\ref{#1})}
\newcommand{\mr}[1]{\mathrm{#1}}
\newcommand{\be}{\begin{equation}}
\newcommand{\ee}{\end{equation}}

\newcommand{\kb}{k_{\mr{B}}}

\newcommand{\figta}{$\left(\mathrm{a}\right)\;$}
\newcommand{\figtb}{$\left(\mathrm{b}\right)\;$}
\newcommand{\figtc}{$\left(\mathrm{c}\right)\;$}
\newcommand{\figtd}{$\left(\mathrm{d}\right)\;$}

\newcommand{\figa}{$\left(\mathrm{a}\right)$}
\newcommand{\figb}{$\left(\mathrm{b}\right)$}
\newcommand{\figc}{$\left(\mathrm{c}\right)$}
\newcommand{\figd}{$\left(\mathrm{d}\right)$}

\newcommand{\gohm}{\;\mr{G}\Omega}
\newcommand{\mohm}{\;\mr{M}\Omega}
\newcommand{\kohm}{\;\mr{k}\Omega}

\newcommand{\mhz}{\;\mr{MHz}}

\newcommand{\mk}{\;\mr{mK}}

\newcommand{\pa}{\;\mr{pA}}
\newcommand{\fa}{\;\mr{fA}}

\newcommand{\mv}{\;\mr{mV}}

\newcommand{\farad}{\;\mr{F}}

\newcommand{\mum}{\;\mu\mr{m}}

\newcommand{\muev}{\;\mu\mr{eV}}
\newcommand{\muv}{\;\mu\mr{V}}
\newcommand{\nm}{\;\mr{nm}}

\newcommand{\permum}{\;\mu\mr{m^{-3}}}
\newcommand{\hz}{\;\mr{Hz}}

\newcommand{\pinj}{P_\mr{inj}}
\newcommand{\ainj}{A_\mr{inj}}
\newcommand{\nqp}{n_\mr{qp}}
\newcommand{\nqpe}{n_\mr{qp}}
\newcommand{\csigma}{C_{\Sigma}}
\newcommand{\alox}{\mr{Al}_2\mr{O}_3}
\newcommand{\nng}{n_\mr{g}}
\newcommand{\cg}{C_\mr{g}}
\newcommand{\tb}{T_{\mr{b}}}
\newcommand{\vb}{V_{\mr{b}}}
\newcommand{\vg}{V_{\mr{g}}}
\newcommand{\ngo}{n_{\mr{g}0}}
\newcommand{\rt}{R_{\mr{T}}}
\newcommand{\rl}{R_0}
\newcommand{\vgo}{V_{\mr{g}0}}
\newcommand{\ag}{A_{\mr{g}}}
\newcommand{\ts}{T_{\mr{S}}}
\newcommand{\tn}{T_{\mr{N}}}
\newcommand{\nqpo}{n_{\mr{qp},0}}
\newcommand{\ec}{E_\mr{C}}
\newcommand{\sigmat}{\sigma_\mr{T}}
\newcommand{\rhon}{\rho_\mr{N}}
\newcommand{\nef}{N(E_{\mr{F}})}
\newcommand{\gammaqp}{\Gamma_{\mr{qp}}}

\begin{document}

\title{Hybrid single-electron turnstiles with thick superconducting electrodes for improved quasiparticle relaxation}

\author{J. T. Peltonen}
\email{joonas.peltonen@aalto.fi}
\affiliation{Department of Applied Physics, Aalto University School of Science, POB 13500, FI-00076 AALTO, Finland}

\author{A. Moisio}
\affiliation{Department of Applied Physics, Aalto University School of Science, POB 13500, FI-00076 AALTO, Finland}

\author{V. F. Maisi}
\affiliation{Department of Applied Physics, Aalto University School of Science, POB 13500, FI-00076 AALTO, Finland}

\author{M. Meschke}
\affiliation{Department of Applied Physics, Aalto University School of Science, POB 13500, FI-00076 AALTO, Finland}

\author{J. S. Tsai}
\affiliation{RIKEN Center for Emergent Matter Science, Wako, Saitama 351-0198, Japan}
\affiliation{Department of Physics, Tokyo University of Science, Kagurazaka, Tokyo 162-8601, Japan}

\author{J. P. Pekola}
\affiliation{Department of Applied Physics, Aalto University School of Science, POB 13500, FI-00076 AALTO, Finland}

\date{\today}

\begin{abstract}
We demonstrate shadow evaporation-based fabrication of high-quality ultrasmall normal metal -- insulator -- superconductor tunnel junctions where the thickness of the superconducting electrode is not limited by the requirement of small junction size. The junctions are formed between a film of manganese-doped aluminium acting as the normal conducting electrode, covered by a thicker, superconducting layer of pure Al. We characterize the junctions by sub-gap current--voltage measurements and charge pumping measurements in a gate-driven hybrid single-electron transistor, operated as a turnstile for single electrons. The technique allows to advance towards turnstiles with close to ideally thermalized superconducting reservoirs, prerequisite for reaching metrological current quantization accuracy in a hybrid turnstile. We further present an alternative way to realize small junctions with thick Al leads based on multi-angle deposition. The work enables the future investigation of turnstiles based on superconductors other than Al, and benefits various other Al tunnel junction devices for which quasiparticle thermalization is essential.
\end{abstract}

\date{\today}

\maketitle

\section{Introduction}

The performance of a multitude of superconducting mesoscopic devices is hindered by both a residual nonequilibrium quasiparticle (qp) density, as well as that arising from the device operation. Crucially, the rapid progress in improving the coherence of various types of superconducting quantum bits has recently enabled the study of the degrading effect of qp excitations~\cite{martinis09,catelani11} on the relaxation and dephasing times in various types of qubits~\cite{lenander11,sun12,wenner13,riste13,pop14,gustavsson16}. Similarly, excess qps limit the quality factors of superconducting resonators~\cite{devisser11,devisser14} and the sensitivity of other Josephson devices such as single-Cooper-pair transistors~\cite{joyez94,aumentado04,ferguson08} and weak links~\cite{levensonfalk14,zgirski11}. Means to reduce the quasiparticle density with the help of qp thermalization by normal-metallic traps~\cite{court08,rajauria09,riwar16} or magnetic vortices~\cite{ullom98,nsanzineza14,wang14,peltonen11,taupin16} have been investigated extensively. In particular, efficient quasiparticle traps~\cite{oneil12,nguyen13,nguyen14} have been found to be vital for the operation of high-power electronic microcoolers based on Normal metal--Insulator--Superconductor (NIS) tunnel junctions~\cite{clark04,lowell13,zhang15}.

In this work we focus on another device based on two ultrasmall NIS junctions in series, in the form of a single-electron transistor (SET) with normal (N) island coupled to superconducting (S) leads, namely a hybrid SINIS turnstile~\cite{pekola08}. Operating such a charge pump under periodic gate voltage drive at frequency $f$ produces a quantized current $I=nef$, where $n$ is an integer and $e$ is the electron charge. At the same time, a power $\pinj\approx\Delta f$ from nonequilibrium qps with energy $E\gtrsim\Delta$, the superconducting energy gap, is injected into each S lead of the turnstile. Overheating of the S electrodes leads to an excess current through the device~\cite{knowles12,pekola13}. Consequently, the accuracy of current quantization deteriorates well before the intrinsic limits due to higher order tunneling process start to play a role~\cite{averin08} if the qps are not efficiently removed from the vicinity of the injecting junctions by optimized qp traps and design of the superconducting leads.

In Ref.~\onlinecite{knowles12} it was first shown experimentally how the accuracy of the pumped current in a SINIS turnstile is deteriorated by excess qps. Besides efficient shielding from stray microwave radiation to reduce the excess qp density to levels $\nqp\ll 0.1\mum^{-3}$\;\cite{pekola10,saira12}, the accuracy depends critically on the geometry of the S leads close to the NIS junctions: A quickly widening lead profile is required for fast diffusion of qps away from the junction area. It was concluded that reaching metrological accuracy as a quantized single-electron current source, corresponding to a reasonable current $I\approx 10\pa$ and relative uncertainty $\delta I/I<10^{-6}$ at realistic driving frequencies of the order of $50\mhz$, requires making the superconducting aluminium leads an order of magnitude thicker than the typical thin film thickness of $20-50\nm$ and minimizing the film resistivity. Simultaneously, the transparency of the contact to the normal metal qp trap needs to be optimized, while retaining consistently high charging energy $\ec=e^2/2\csigma\gtrsim 2\Delta$ ($\csigma$ stands for the total capacitance of the turnstile island) in combination with good control over the tunnel resistances of the ultrasmall junctions. These requirements present a grand challenge for conventional NIS junction fabrication by means of shadow evaporation: Keeping a small and well-defined junction overlap area constrains the thickness of the Al leads (deposited as the first film) as the normal metal needs reliable step coverage of the Al layer.

\begin{figure}[htb]
\includegraphics[width=\columnwidth]{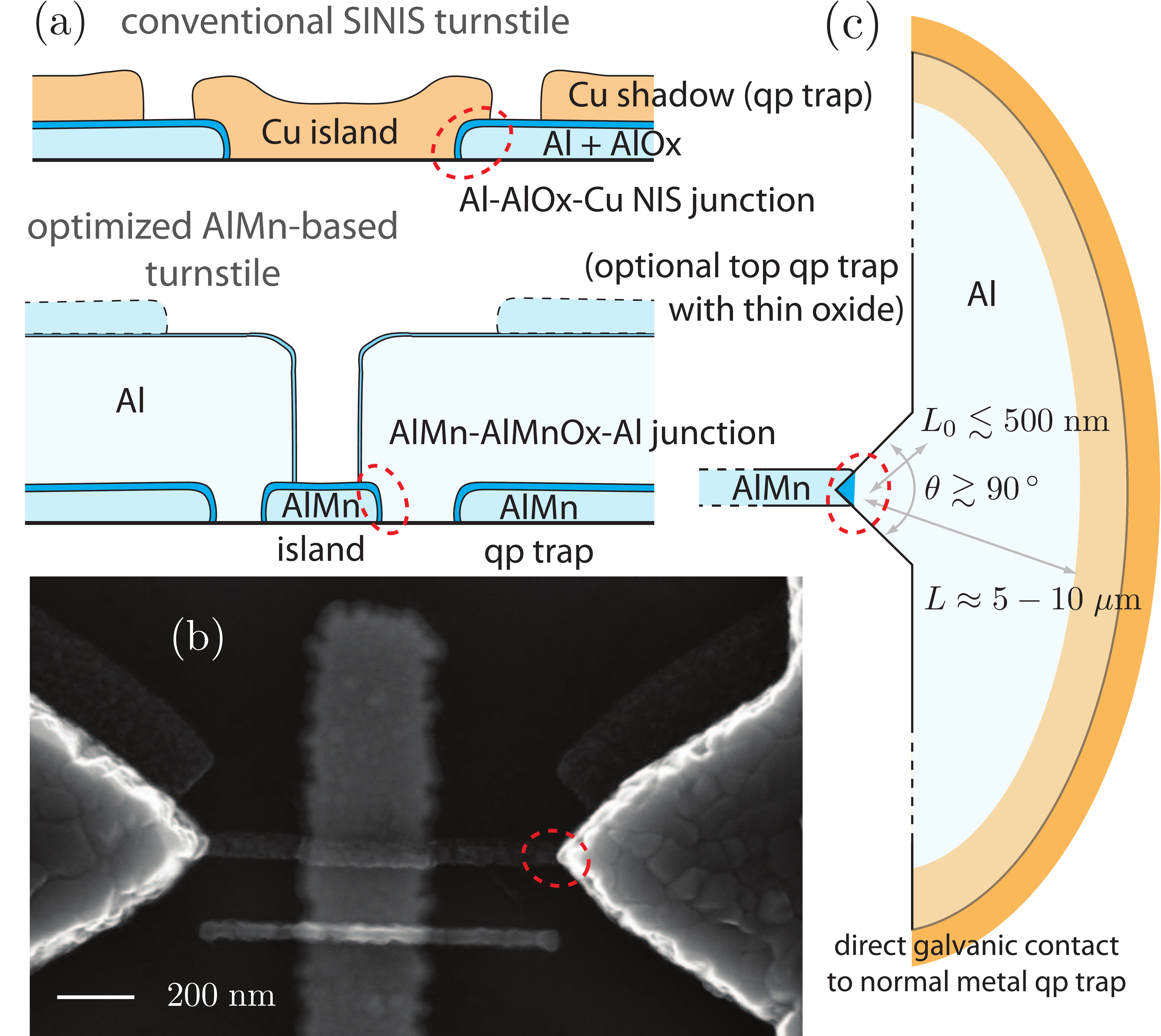}
\caption{(color online) \figta Schematic cross section of a conventional SINIS turnstile (top), and a turnstile with the N and S electrodes deposited in opposite order, thus enabling thick superconducting leads (bottom). The red dashed line points out how the turnstile junctions are formed at the overlaps of the N and S electrodes. \figtb Scanning electron micrograph of a typical AlMn-based device, showing the turnstile N island and the junctions in an enlarged view. The single gate for combined dc offset and rf drive (the vertical, approximately $200\nm$ wide strip) is only capacitively coupled to the island, separated from it by an approximately $50\nm$ thick $\alox$ dielectric grown by ALD. The gate is omitted from the cross section. \figtc Sketch of the superconducting lead geometry of the investigated AlMn samples, emphasizing the wide opening angle $\theta$ to enhance quasiparticle diffusion away from the junctions. The AlMn qp trap below the Al electrode is omitted for clarity.} \label{fig:sample}
\end{figure}

In this work we demonstrate a straightforward fabrication technique employing standard two angle shadow evaporation, yielding reproducible tunnel junctions with down to $20\times20\nm^2$ well-defined overlap area combined with up to $\mum$-thick superconducting leads. The thickness of the S electrodes is limited only by the height of the sacrificial resist. We further show that it appears to be a feasible approach to meet all the above-mentioned criteria, required when aiming for metrological accuracy in a metallic SINIS turnstile. The process is based on first forming the normal metal island of the turnstile out of aluminium lightly doped with manganese impurities, cf. Fig.~\ref{fig:sample}~\cite{clark04}. The AlMn film can be oxidized in-situ, leading to tunnel barriers practically identical in quality to the oxide on top of undoped aluminium, and the superconducting leads are subsequently deposited in the last step, allowing for very large film thickness if desired. Here we present basic low-frequency electrical characterization of such an AlMn turnstile, together with initial charge pumping experiments. Crucially, we show that the junctions are of promising quality, despite the accuracy of the room-temperature current measurement does not yet show if metrological turnstile accuracy would ultimately be limited by the junction leakage current. We further demonstrate an alternative way to realize turnstiles with thick S reservoirs, based on spatial variation of the Al thickness close to the junction resulting from multi-angle Al deposition.

\section{Samples with thick leads}

The basic idea is illustrated by Fig.~\ref{fig:sample}~\figa, comparing the simplified cross section of an AlMn-based turnstile (bottom) to a regular device with Al--AlOx--Cu junctions. All prior experiments on the hybrid turnstiles have been done on conventional devices sketched at the top, due to the favorable properties of the native Al oxide forming the tunnel barriers. They are created by two-angle shadow evaporation of pure aluminium through a suspended resist, followed by in-situ oxidation before finally depositing the N island using the same mask but another tilt angle with respect to the evaporation source, typically with Cu or AuPd as the metal of choice. In such a process where the superconducting electrodes are deposited as the first layer, their thickness is necessarily limited, leading to ineffective qp thermalization. Main factors requiring moderate Al thickness include the need for high $\ec$, i.e., small and well-controlled junction overlap area, as well as issues with self-blocking of the shadow mask due to the first deposited film. In this work we characterize turnstiles based on the design shown in the bottom half of panel~\figa, utilizing dilute (typically of the order of 1\%) Mn doping to fully suppress the superconductivity in aluminium while retaining virtually identical properties of the tunnel barrier. As a result, the superconducting Al leads of the turnstile can be deposited as the second film, on top of the oxidized normal-conducting AlMn island. The reversed deposition order allows for a very large difference in the N island and S lead thicknesses while keeping the junction area well defined, and evaporating the Al layer at zero tilt minimizes problems from mask blocking. Note that evaporation from an AlMn alloy target will enrich the Mn in the final film as the vapor pressure at a given temperature is about one order of magnitude higher for Mn in comparison with Al.

Figure~\ref{fig:sample}~\figtb shows a scanning electron micrograph of a typical AlMn turnstile. Here, the more than an order of magnitude difference in the S and N film thicknesses is very evident in the image contrast. As found out in Refs.~\onlinecite{knowles12} and~\onlinecite{pekola13}, besides large thickness it is essential that the leads have a wide opening angle $\theta$ at the junction, ideally opening to half space very close to the junction region. We approximate this in the AlMn devices with the S electrode geometry sketched in Fig.~\ref{fig:sample}~\figc. Before fully widening out, a gradual opening of the leads up to a short distance $L_0\lesssim 0.5\mum$ is needed to prevent the suspended shadow mask between the two electrodes from collapsing. The quick opening of the Al lead at $L_0$ becomes just visible also at the right edge of Fig.~\ref{fig:sample}~\figb.

To further enhance qp thermalization, the turnstiles feature a normal metal (AuPd) qp trap in direct contact with the S electrodes as illustrated in Fig.~\ref{fig:sample}~\figc. In our samples the N region starts at a relatively short distance $L\approx 5-10\mum$ away from the turnstile junctions, a compromise between efficient qp trapping and leakage current due to the inverse proximity effect. Moreover, in Fig.~\ref{fig:sample}~\figta it is emphasized that for the conventional (inverted) process an intrinsic large-area, normal metal qp trap is always formed by N shadow copy of the S lead pattern on top of (underneath) the Al layer. This shadow trap is separated from the Al layer by an oxide barrier formed alongside the relatively opaque turnstile junctions. Importantly, in AlMn-based turnstiles, another optimized, oxide-separated shadow trap can be deposited on top of the thick S leads at a third evaporation angle. For this separate trap layer the tunnel barrier transparency can be controlled independently from the turnstile junctions and optimized to achieve efficient qp evacuation close to the junctions. Note that an AlMn-based efficient trap with transparent oxide can be combined also with the regular turnstile process when it is deposited as the first layer, located beneath the Al leads~\cite{nguyen13,nguyen14}.

Owing to advantages of forming tunnel junctions to an oxidized normal metal, AlMn has been investigated extensively for the realization of high-power SINIS coolers with a thin N electrode combined to massive, well-thermalized and low-resistive Al leads~\cite{clark04,oneil12,lowell13,zhang15,nguyen13}. In these devices where efficient qp relaxation is essential for high cooling power, the junctions with overlap areas up to several hundred $\mu\mr{m}^2$ are typically realized using a two-step fabrication process, combined with ion milling of the AlMn surface before oxidation and deposition of the S electrode. Smaller AlMn junctions with sub-micron dimensions, typically deposited by two-angle shadow evaporation, have been fabricated as well~\cite{taskinen06,karvonen07,underwood11,ciccarelli16,perez14,enrico16,fornieri16}. However, they have typically not been measured in highly shielded setups which would allow probing the intrinsic sub-gap leakage currents for assessing the junction quality for metrological purposes. Here we consider ultrasmall AlMn-based NIS junctions under such conditions.

We wish to state here briefly the results of the qp diffusion model of Refs.~\onlinecite{knowles12} and~\onlinecite{pekola13}, applied to the optimized S lead geometry in Fig.~\ref{fig:sample}~\figc. Solving a 1D diffusion equation
\be
\nabla^2\nqp=\frac{\nqp-\nqpo}{\lambda^2}\label{diffeq}
\ee
in polar coordinates, for a device with a direct contact qp trap at distance $L$ away from the junction the total qp density at the junction is given by
\be
\nqpe=\nqpo+\frac{\lambda\pinj}{D\ainj}\frac{I_0(\frac{L}{\lambda})K_0(\frac{r_0}{\lambda})-I_0(\frac{r_0}{\lambda})K_0(\frac{L}{\lambda})}{I_0(\frac{L}{\lambda})K_1(\frac{r_0}{\lambda})+I_1(\frac{r_0}{\lambda})K_0(\frac{L}{\lambda})}.\label{nqp1}
\ee
Here, $\nqpo$ stands for the equilibrium qp density at the direct contact trap (assumed to be thermalized to the cryostat bath temperature), whereas $I_{\alpha}(x)$ and $K_{\alpha}(x)$ are modified Bessel functions. Further, $\ainj=\theta r_0 d$ is the area into which the power $\pinj$ is injected at the junction, assumed to be located at distance $r_0$ in the approximately sector-shaped leads (thickness $d$, opening angle $\theta$). The quantity $\lambda$ can be interpreted as a qp relaxation length, defined via $\lambda^2=d\sqrt{2(\kb\ts/\Delta)}/(\sqrt{\pi}\sigmat\rhon)$. Here, $\ts$ is the (effective) temperature of the superconductor, determined self-consistently by equating the total qp density to an equilibrium density at $\ts$. $\rhon$ denotes the normal state resistivity of the S electrode, $\sigmat$ is the transparency (specific conductance) of the oxide barrier qp trap covering the electrode, and $D=\lambda^2\Delta/(e^2\rhon\nef)$ can be interpreted as a diffusion constant. Finally, $\nef$ denotes the normal state density of states in the S leads at the Fermi energy. For most realistic values of $\rhon$ and $\sigmat$ for a device without an optimized, oxide-separated qp trap, we find $r_0\ll L\ll\lambda$. Explicitly, assuming $d=100\nm$, $\ts=120\mk$, $\Delta=200\muev$, $\rhon=10^{-8}\;\Omega\mr{m}$, and $\sigmat^{-1}=1\;\mr{k}\Omega\mu\mr{m}^2$ gives $\lambda\approx 43\mum$, whereas $\sigmat^{-1}=100\;\Omega\mu\mr{m}^2$ corresponds to $L\approx 13\mum$. In the limit $r_0\ll L\ll\lambda$ Eq.~\equref{nqp1} yields to a good approximation
\be
\nqpe\approx\nqpo+\frac{\pinj}{D\ainj} r_0\ln(L/r_0).\label{nqp2}
\ee
This result is to be contrasted with $\nqpe\approx\nqpo+\pinj L/(D\ainj)$ obtained for S leads with constant cross section $\ainj$, demonstrating a reduction in the excess qp density due to quickly broadening electrodes by a factor $(r_0/L)\ln(L/r_0)\ll 1$. Assuming $\theta=\pi/2$, $L=10\mum$, $r_0=20\nm$, $\sigmat^{-1}=1\;\mr{k}\Omega\mu\mr{m}^2$, $I=10\pa$ (at $f=I/e\approx 62\mhz$) and keeping the other parameters at the above values, we find a qp density $\nqpe\approx 0.5\mum^{-3}$ at the junction. For an aluminium device with $\nef=1.45\times 10^{47}\;\mr{J}^{-1}\mr{m}^{-3}$ and total normal state resistance $\rt=2\mohm$ ($1\mohm$ per junction), this corresponds to a residual tunneling rate~\cite{pekola13} $\gammaqp\approx\nqpe/(e^2\rt\nef)\approx68\hz$, and hence an approximate relative accuracy $1\times 10^{-6}$.

Except for the junction deposition in the final step, the sample fabrication process is identical to the one described in Ref.~\onlinecite{peltonen17} in more detail: In a first round of electron beam lithography and metal deposition, the turnstile gate and a large-area ground plane electrode for on-chip filtering are formed from a $30\nm$ thick layer of Au, subsequently covered by $50\nm$ insulating $\alox$. A second lithography step defines the normal metal traps ($30\nm$ of AuPd), starting at $L\approx 5-10\mum$ away from the junction region and extending into bonding pads. In the third and final lithography step, a suspended Ge mask with the turnstile island and lead pattern is prepared. The junctions are then formed by evaporation of $30\nm$ AlMn at a finite tilt angle, followed by in-situ oxidation and the deposition of the thick Al layer normal to the substrate. As the source of the manganese-doped aluminium, we use Goodfellow AlMn with nominal Mn concentration of 0.3\%, evaporated from a Fabmate graphite crucible. A new target was prepared for each deposition. With a $400\nm$ thick P(MMA-MAA) copolymer sacrificial layer under the Ge mask, we evaporated up to $350\nm$ Al. However, the thickness of the resist stack can straightforwardly be increased beyond $1\mum$, allowing for small tunnel junctions with micron-thick Al leads.

\section{Low-frequency electrical characterization}

We now turn to discussion of low-frequency, non-driven electrical characterization of the SETs. All measurements were performed on a double-shielded, microwave-tight sample stage~\cite{saira12,knowles12} to limit the residual qp density due to microwave radiation from higher temperature stages in the cryostat. The on-chip ground plane further assists in the filtering. The sample box was thermalized to the mixing chamber of a dilution refrigerator, and measurements were done at the base temperature $\tb\approx 50\mk$. At the lowest temperature stage the measurement lines were filtered by Thermocoax cable~\cite{zorin95}, including the gate line where a shorter Thermocoax section was used to retain the bandwidth necessary for driven turnstile operation.

\begin{figure}[htb]
\includegraphics[width=\columnwidth]{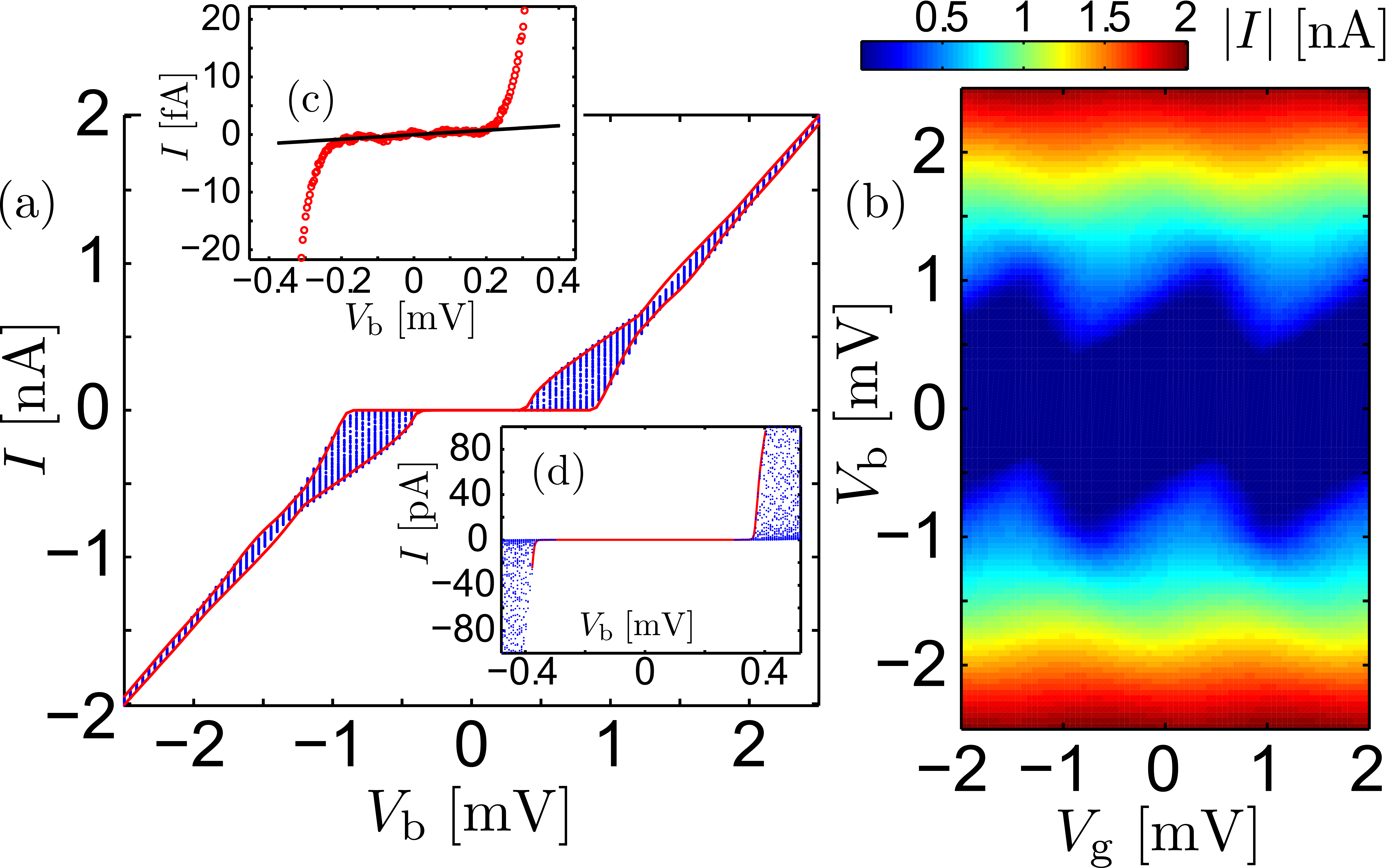}
\caption{(color online) \figta Current--voltage characteristics of an AlMn device, measured at the base temperature $\tb\approx 50\mk$. At each bias voltage $\vb$, the gate voltage $\vg$ is swept over a range corresponding to several charge states on the island. The solid lines are obtained from a standard steady-state master equation with golden rule-based rates describing sequential tunneling of single electrons. \figtb Stability diagram corresponding to the envelope plot in~\figa. \figtc Sub-gap leakage current in the gate open state (maximum current). The red dots show the average of N repeated bias voltage sweeps during which the gate position was compensated to stay at the maximum current. The black solid line is a linear fit using the points with $|\vb|<100\muv$, yielding a slope $\eta=\rt/\rl\approx 3.5\times 10^{-6}$. \figtd The red circles are same as in panel \figtc but on a larger scale. The small blue symbols correspond to a similar envelope measurement as in panel \figta performed at a higher gain of the current preamplifier, indicating that the gate open position is tracked accurately as seen by the course of the red line with respect to these points. \label{fig:dc}}
\end{figure}

The blue dots in Fig.~\ref{fig:dc}~\figta show a typical envelope plot of the AlMn turnstile, i.e., the current at each bias voltage $\vb$, measured during a sweep of the gate voltage $\vg$ over several multiples of the gate charge $\nng=\cg\vg/e$. The same data is plotted in Fig.~\ref{fig:dc}~\figtb as a function of both $\vb$ and $\vg$. This stability diagram of the hybrid SET clearly features the Coulomb diamonds of blocked current flow at low bias. As a central result of this work, the measurement clearly demonstrates the reliable fabrication of small junctions with high charging energy, coupled to large Al reservoirs. The red solid lines in panel~\figta show the calculated maximum and minimum current as a function of $\vb$. For a symmetric SET, these ``gate open'' and ``gate closed'' states with maximum current and maximum blockade correspond to $\nng=1/2$ and $\nng=0$, respectively.

For theoretical modeling of the measurements in Fig.~\ref{fig:dc} we use a standard master equation approach to describe the occupation probabilities of the different charge states on the SET island, see, e.g., Ref.~\onlinecite{pekola13} and references therein. The tunneling rates entering the calculation follow the golden rule-based orthodox theory of single-electron tunneling~\cite{averin91}. Self-heating or cooling of the N island by the applied bias is for simplicity approximated by assuming standard $T^5$ electron--phonon coupling in the AlMn film with a coupling constant $\Sigma\approx 2.5\;\mr{nW}\mu\mr{m}3{-3}K^{-5}$ as for Cu. Performing fits of the model to the measured stability diagram, we find a total normal-state resistance $\rt\approx 925\kohm$, Al superconducting energy gap $\Delta\approx 200\muev$, charging energy $\ec\approx 320\muev$, and a junction asymmetry $R_{\mr{T,L}}/R_{\mr{T,R}}=3$. Note that based on the period of the gate modulation $\delta\vg\approx 1.9\mv$, we infer the relatively strong gate coupling $\cg\approx 8\times 10^{-17}\farad$, vs. $\csigma\approx 2.5\times 10^{-16}\farad$. The gate capacitance can be straightforwardly moderately reduced in future devices to reach $\ec\approx 2\Delta$, an optimal value considering both high turnstile operation speed and suppression of higher order tunneling processes~\cite{averin08}.


The red circles in Fig.~\ref{fig:dc}~\figtc show a measurement of the sub-gap IV characteristic in the gate open state. To improve the signal-to-noise ratio, the results of $x$ repeated bias voltage sweeps are averaged. The same data (red line) is included in panel~\figtd on a larger scale, together with an envelope measurement similar to panel~\figa, indicating that the $\vg$ value corresponding to gate open state is followed closely. From a linear fit to the sub-gap current in Fig.~\ref{fig:dc}~\figtc (solid black line) we find an estimate of the gate-open zero bias resistance $\rl\approx 270\gohm$. This value corresponds to a leakage parameter~\cite{dynes84,pekola10} $\eta=\rt/\rl\approx 3.5\times 10^{-6}$, typical among the investigated hybrid AlMn SETs and on a similar level with our conventional Al--Cu devices. The low levels of sub-gap leakage current make the junctions fabricated with this technique promising for further investigation as accurate single-electron turnstiles, or in several other applications such as low-temperature electronic thermometers and coolers. We note that the value of $\rl$ may already be affected by parasitic leakage resistances in the measurement circuit, and a more stringent test of the sub-gap leakage current would be achievable with single junctions with lower $\rt$, or in an electron counting experiment~\cite{saira12,peltonen17}.

\section{Pumping measurements}

\begin{figure}[htb]
\includegraphics[width=\columnwidth]{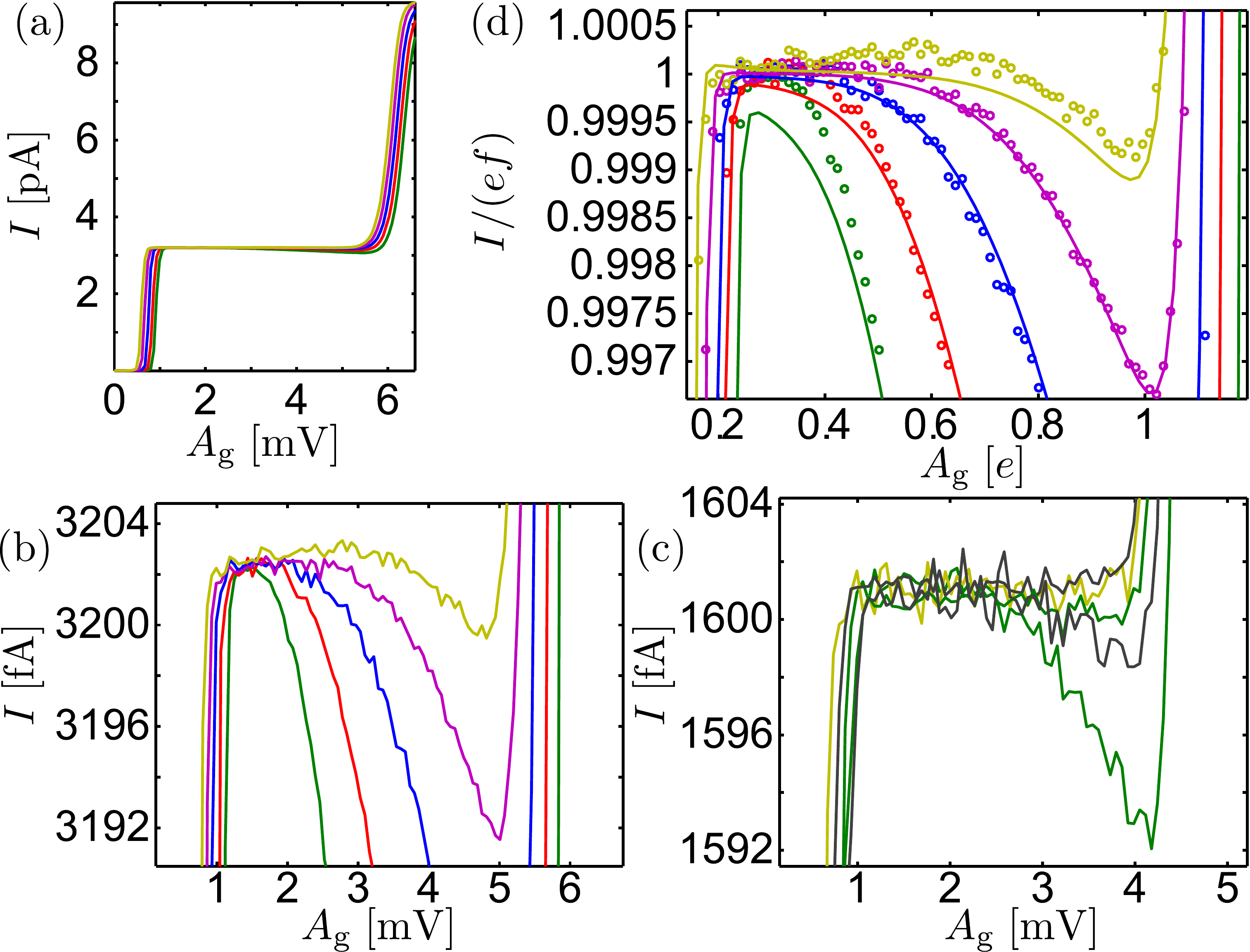}
\caption{(color online) \figta Average pumped current for the AlMn turnstile under continuous sinusoidal gate drive at $f=20\mhz$. The different curves are obtained by changing the bias voltage $\vb$ from $140\muv$ to $220\muv$ in steps of $20\muv$. \figtb Zoomed-in view of the pumping plateaus at $f=20\mhz$ in panel~\figa, and \figtc the result of a similar measurement at $f=10\mhz$. \figtd Data in panel \figtb normalized to $ef$. The solid lines are calculated assuming $\tn=\ts=160\mk$. \label{fig:pump}}
\end{figure}

We next discuss measurements of charge pumping in an AlMn turnstile under continuous sinusoidal gate drive $\vg(t)=\vgo+\ag\sin(2\pi ft)$. Figure~\ref{fig:pump}~\figta shows typical behavior of the average pumped current $I$ when the amplitude $\ag$ of the drive at fixed frequency $f=20\mhz$ is increased. The dc offset $\vgo$ is fixed to a value corresponding to the gate open state. As expected, we observe a flat plateau at $I=ef\approx 3.204\pa$, and the onset to the second quantized step at $I=3ef$. Figure~\ref{fig:pump}~\figtb displays an enlarged view of the first plateau in panel~\figa, whereas panel~\figtc shows the result of a similar pumping measurement at the lower frequency $f=10\mhz$. As one of the main results of this work, the AlMn device with small tunnel junctions and large Al reservoirs demonstrates promising pumping performance.

Each curve corresponds to a fixed bias voltage $\vb$, ranging from $140\muv$ to $220\muv$ with $20\muv$ step (with onset of $I=ef$ plateau at lowest to highest $\ag$, respectively) around the optimal value $\vb\approx\Delta/e$\;~\cite{pekola08}. Pumping errors due to backtunneling qps~\cite{kemppinen09,pekola13} are evident in both panels~\figtb and~\figc. They manifest as the decrease of the current below $I=ef$ with increasing $\ag$, before the climb to the second plateau at $I=3ef$ just visible in panel~\figta at the largest $\ag$. The dip in the pumped current gets more pronounced towards lower values of $\vb$ and higher drive frequencies. In particular for our device with relatively high $\rt$ and $\ec$, backtunneling is the main factor limiting the range of $\ag$ and $\vb$ for which a flat plateau is obtained.

The open circles in Fig.~\ref{fig:pump}~\figtd show the data of panel~\figtb normalized to $ef$, after taking into account corrections to the nominal amplifier gain. We find that the plateaus at $I=ef$ are flat against variations in $\ag$ and $\vb$ down to the level $10^{-4}$. This relative accuracy is close to the best reported result for an Al--Cu SINIS turnstile, achieved in Ref.~\onlinecite{knowles12} in a device with quickly widening Al electrodes of regular thickness. The solid lines show the result of simulations, obtained by finding a periodic steady-state to the master equation~\cite{pekola13} using parameters estimated from the independent dc measurements in Fig.~\ref{fig:dc}. In the calculation, we assume constant $\tn$ independent of the drive amplitude, finding reasonable agreement on the first pumping plateau at $\tn=160\mk$. For simplicity, the S leads are also described by $\ts=160\mk$, corresponding to a qp density $\nqp\approx 1\permum$. Note that for our initial AlMn device the estimation of $\nqp$ is subject to a significant uncertainty, and determining its increase with the drive frequency cannot be performed reliably for the data in Fig.~\ref{fig:pump}. Compared to the best thermalized sample investigated in Ref.~\onlinecite{knowles12}, the AlMn turnstile has an order of magnitude larger $\rt$, leading to a correspondingly reduced effect from $\nqp$: the excess current scales inversely with the device resistance. We note that similar to the sub-gap leakage discussed above, a stringent limit to $\ts$ can be obtained with devices with more transparent tunnel junctions.

In the present setup, the accuracy is limited by the basic, non-traceable measurement scheme based on direct current measurements with a conventional room-temperature transimpedance amplifier (Femto DDPCA-300 or Femto LCA-2-10T). This is in contrast to Ref.~\onlinecite{knowles12}, where the turnstile current was compared to that produced by voltage applied from a calibrated voltage source over a calibrated resistor. In our experiment, the amplifier gain is calibrated down to the level of $10^{-4}$. More crucially, the pumping measurements in Fig.~\ref{fig:pump} are limited by drifts of the amplifier offset voltage and current. Under optimized conditions, the basic scheme has a typical peak-to-peak noise level of $5-10\fa$ when measuring the $I=ef$ pumping plateaus at $10-20\mhz$. Each curve in Fig.~\ref{fig:pump} is the average of typically 10--15 amplitude sweeps. Between each such scan, the gate offset $\ngo$ is re-checked to remain at the prescribed working point (at $\ngo=0.5$ for a symmetric SET) by scanning $\ngo$ at a fixed $\ag$ and finding the center of the resulting current peak. Curves where an obvious offset charge jump has happened are omitted from the averaging. After this procedure, it is possible to achieve the approximately $0.5\fa$ noise level seen in panels~\figb--\figd.

\section{Discussion}

\begin{figure}[htb]
\includegraphics[width=0.8\columnwidth]{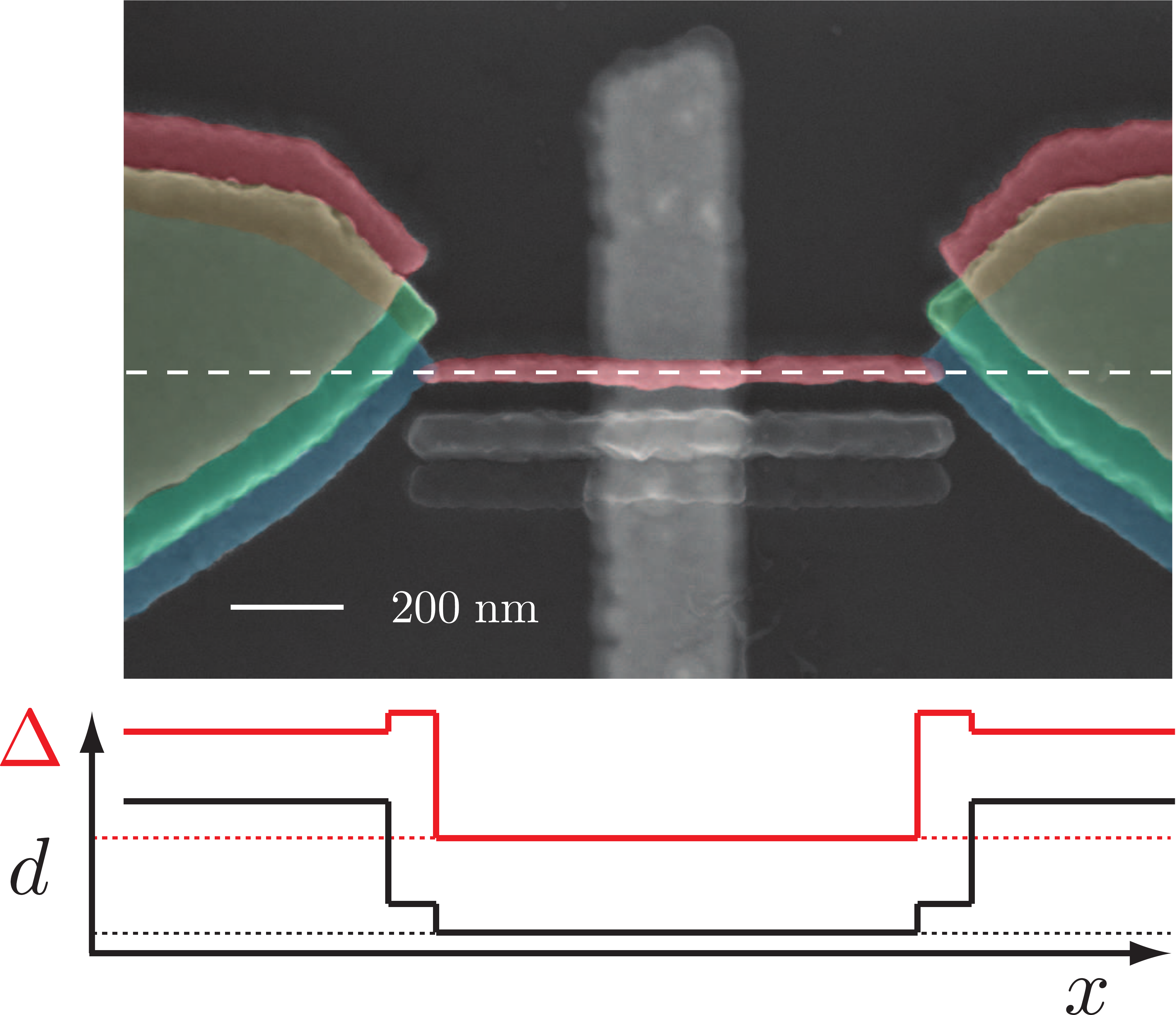}
\caption{(color online) Scanning electron micrograph of a ``gap-engineered'' turnstile with superconducting leads featuring a two-step thickness profile close to the NIS junctions. In this Al--Cu device, blocking of the shadow mask opening for the N island by the Al deposition needs to be taken into account in the pattern design, and limits the ultimate Al thickness. The same principle can be applied to AlMn-based structures. The black (red) line shows a sketch of the Al thickness (superconducting energy gap) profile along the white dashed line. \label{fig:gapengineering}}
\end{figure}

Compared to the regular Al-first turnstile fabrication process, the AlMn-based scheme has several attractive features. Along with increasing the thickness of the S leads, their normal state resistivity is notably reduced towards the bulk resistivity as surface scattering is reduced for thicker films. Interestingly, also the Al superconducting gap approaches the bulk value $\Delta\approx 180\muv$, vs. $\Delta\approx 220-230\muv$ often found in turnstiles with thin leads. Forming the tunnel barrier on the normal AlMn island allows to investigate turnstiles with other superconductors besides Al, both with larger (e.g., Nb, V) or smaller (Ti) gap, whose native oxides are poorly suited to create high quality NIS junctions. What is more, the thick Al leads can be deposited at two different angles with different thicknesses so that the gap is larger right at the junction, and reduces close to the bulk value after approximately $100\nm$ further away. Figure~\ref{fig:gapengineering} illustrates an example of such a ``gap engineered'' turnstile following this principle. Besides obtaining the desired thick leads, the resulting gap profile further assists in the qp evacuation away from the junction~\cite{ferguson06,sun12}. In particular, such thickness-based control of Al energy gap, boosted by oxygen exposure during the deposition~\cite{naaman06}, has been studied to improve the performance of all-superconducting single-Cooper-pair transistors~\cite{joyez94,aumentado04,ferguson06} as well as superconducting qubits~\cite{sun12}.

In contrast to several common types of Al-based NIS junctions, e.g., Al--Cu, Al--Ag, or Al--AuPd, in Ref.~\onlinecite{julin16} it was found out to be possible to anneal the AlMn-based devices similar to purely Al SIS junctions~\cite{julin10}. Annealing studies could shed further light onto the issue of small effective transmission channel size of the conventional NIS junctions~\cite{maisi11}. The tunnel barrier transparency of our AlMn devices can be further optimized. Already in the present samples the tunnel resistances are sufficiently high when it comes to suppressing higher order error processes~\cite{averin08}. We think the junction quality can still be improved by making the depositions in a cleaner, dedicated evaporation tool, and having more precise control over the Mn concentration. Nevertheless, already the present method, based on preparing a new AlMn target for each deposition, results in junctions with very low sub-gap leakage currents. In earlier work on large-area SINIS coolers, Mn concentration tuning has been typically achieved by co-sputtering Al and Mn from two different targets~\cite{clark04,lowell13,zhang15}. A significant amount of the research has focused on the controlled suppression of the gap while retaining low sub-gap density of states~\cite{ruggiero03,oneil08,oneil10}. For SINIS turnstiles the superconductor density of states is not an issue as we employ pure Al as the counterelectrode and keep AlMn intentionally fully in the normal state.

Manganese-doped aluminium is not the only possible candidate for fabricating turnstiles where the normal island is deposited first. Other materials with suitable natural oxide, e.g. scandium, appear possible as well~\cite{lehtinen16}. We further note that the $T^6$ electron-phonon heat flow in AlMn (i.e., weaker at very low temperatures) compared to $T^5$ in Cu~\cite{taskinen06,underwood11} can be useful, and the island may even be cooled intrinsically~\cite{kafanov09,ciccarelli16}. Finally, active quasiparticle cooling~\cite{ferguson08}, based on qp tunneling in a voltage-biased tunnel junction between two superconductors with unequal gaps~\cite{manninen99}, remains an interesting unexplored option for SINIS turnstiles. It can be combined in a straightforward way with AlMn turnstiles where the gap is slightly reduced in the bulk electrodes.

In summary, we have investigated the fabrication of ultrasmall, low-leakage hybrid tunnel junctions with thick superconducting leads. We have shown that AlMn, proven with NIS coolers, can offer a simple and promising way to realize hybrid SINIS turnstiles with optimized superconducting electrodes. Combined with state-of-the-art high-resolution electron beam lithography and a Ge-based hard mask to obtain consistently high charging energies and reproducible junction parameters, it offers a feasible path towards utilizing hybrid turnstiles as a source of quantized current for metrological applications. Future work is to assess the ultimate accuracy of the thick-leaded turnstiles, initially by a null measurement against a calibrated resistor and a voltage source, followed by on-chip counting of errors in pumping.

{\it Acknowledgments.}
We acknowledge the provision of facilities by Aalto University at OtaNano -- Micronova Nanofabrication Centre. We thank J. S. Lehtinen, E. Mykk\"anen, A. Kemppinen, A. Manninen, I. M. Khaymovich, G. Catelani, A. Hosseinkhani, and Yu. A. Pashkin for useful discussions. We acknowledge financial support from the Academy of Finland (Projects 272218, 284594, and 275167).

\end{document}